\documentclass[useAMS,usenatbib,usegraphicx]{mn2e}
\usepackage{aas_macros}
\usepackage{amssymb}

\title[Spin alignment of haloes around voids]{Spin alignment of dark matter haloes in the shells of the largest voids}
\author[A.~J. Cuesta et al.]{Antonio J. Cuesta,$^1$\thanks{E-mail: ajcv@iaa.es} Juan E. Betancort-Rijo,$^{2,3}$ Stefan Gottl\"ober,$^4$ \newauthor Santiago G. Patiri,$^2$\thanks{Current Address: Department of Astronomy, Case Western Reserve University, 10900 Euclid Ave., 44106 Cleveland, Ohio, USA} Gustavo Yepes$^5$ and Francisco Prada$^1$\\
$^1$Instituto de Astrof{\'\i}sica de Andaluc{\'\i}a (CSIC),
Camino Bajo de Hu\'etor 50, E-18008 Granada, Spain \\
$^2$Instituto de Astrof{\'\i}sica de Canarias,
C/V{\'\i}a L\'actea s/n, E-38200 La Laguna, Tenerife, Spain \\
$^3$Universidad de La Laguna, Departamento de Astrof{\'\i}sica, 
Avda. Astrof{\'\i}sico Fco. S\'anchez s/n, E-38200 La Laguna, Tenerife, Spain \\
$^4$Astrophysikalisches Institut Potsdam,
An der Sternwarte 16, D-14482 Potsdam, Germany \\
$^5$Universidad Aut\'onoma de Madrid, 
Grupo de Astrof{\'\i}sica, E-28049 Madrid, Spain
}

\begin{document}

\bibliographystyle{mn2e}
\maketitle

\begin{abstract}
Using the high resolution cosmological $N$-body simulation MareNostrum Universe we study the orientation of shape and angular momentum of galaxy-size dark matter haloes around large voids. We find that haloes located on the shells of the largest cosmic voids have angular momenta that tend to be preferentially perpendicular to the direction that joins the centre of the halo and the centre of the void. This alignment has been found in spiral galaxies around voids using galaxy redshift surveys. We measure for the first time the strength of this alignment, showing how it falls off with increasing distance to the centre of the void. We also confirm the correlation between the intensity of this alignment and the halo mass. The analysis of the orientation of the halo main axes confirms the results of previous works. Moreover, we find a similar alignment for the baryonic matter inside dark matter haloes, which is much stronger in their inner parts.
\end{abstract}


\section{Introduction}

Understanding how the angular momentum is distributed over the collapsed structures in the Universe and its connection with the large-scale structure distribution of galaxies at present is a key issue to determine its origin (see \citealt{2003astro.ph.12547P}, for a review). Furthermore, angular momentum is fundamental in order to explain the structure and dynamics of galaxy discs and hence it is a basic ingredient to determine their evolution and morphology. The currently accepted paradigm for explaining the origin of the angular momentum is the Tidal Torque Theory \citep{1969ApJ...155..393P,1970Afz...6...581W,1984ApJ...286...38W}, hereafter TTT. In this framework, the angular momentum is transferred to the protohalo by the coupling of the quadrupolar inertia tensor and the tidal tensor during the linear regime of growth of density fluctuations. According to TTT, the epoch of turnaround marks a major change in the amount of angular momentum transferred, as this tidal torquing becomes inefficient due to the small size of the halo after it collapses. However the theory breaks at this epoch as non-linear effects are expected to turn up together with merger and accretion processes. In particular, this causes the dilution of the TTT prediction regarding the spin direction \citep{2002MNRAS.332..325P,2004MNRAS.348..921P}.

None the less, recent observational studies have found a preferential orientation of galaxy discs with respect to their surrounding large-scale structures \citep{2004ApJ...613L..41N,2006ApJ...640L.111T}. Such an alignment is thus expected to be a relic of the alignment already present before the epoch of the turnaround. These works differ both in the methodology and the sample used. For instance, \citet{2004ApJ...613L..41N} used catalogues of nearby galaxies and found an excess of edge-on spiral galaxies highly inclined relative to the supergalactic plane. However this result has two major drawbacks: the sample is limited to galaxies in the vicinity of the Milky Way, which is not very representative of the overall large-scale structure; and the orientation of the large-scale sheetlike distributions like the supergalactic plane is blurred due to redshift distortion at high distances (which can only be determined by redshift). On the other hand, \citet{2006ApJ...640L.111T} used the largest galaxy redshift surveys up to date (2dFGRS and SDSS) to get the sample, and searched for large voids in these surveys to identify spatial orientations not relying on the detection of planes. They found that spiral galaxies located on the shells of the largest cosmic voids have rotation axes that lie preferentially on the surface of the void. As this paper has been written, it has also been found a correlation between the orientation of galaxy discs and local tidal shears using the Two Mass Redshift Survey \citep{2007arXiv0706.1412L}.

In addition, $N$-body cosmological simulations have nowadays enough resolution to study whether dark matter haloes also show this alignment. This result is expected as it is commonly assumed that both dark and baryonic matter shared a similar evolution during early epochs and gained the same specific angular momentum before the formation of the disc (e.g. \citealt{1980MNRAS.193..189F}). However, even in the case of a positive signal, it is expected to be very weak as dark matter haloes are more affected by torques from neighbouring haloes than baryonic matter concentrated in discs. This has recently lead to numerous studies with a variety of methods to select a homogeneous sample of dark matter haloes with the aim of searching a counterpart of the observational results. For example, in \citet{2007ApJ...655L...5A} a multiscale filter is developed in order to distinguish haloes belonging to filaments or walls. This code infers the morphology of the structure surrounding a halo by calculating the relations between the eigenvalues of the hessian of the density field. In a dynamical approach, \citet{2007MNRAS.375..489H} use the number of positive eigenvalues of the hessian of the gravitational potential to classify the environment in which a halo resides. Both papers conclude that haloes in walls have spin vectors that tend to lie in the plane of the host wall, but haloes in filaments show only a weak trend for their angular momenta to be aligned with the filament direction. However, the method proposed by \citet{2006ApJ...640L.111T} provides a very clean way to characterize the orientation of the large-scale distribution (i.e. the reference direction to measure any kind of alignment) by using the centre of a large void. This method, besides its robustness in observations with respect to redshift distortion, owes its convenience to the well known fact that the direction of maximum compression (that of the smallest eigenvalue of the deformation tensor) turns out to be very well correlated with the direction to the centre of the void.

The study of this type of alignments is specially interesting in the context of gravitational lensing. A key assumption of this method is that the observed galaxy ellipticity correlations come only from the distortion caused by the gravitational shear, with all the intrinsic terms being negligible (e.g. \citealt{2006MNRAS.371..750H}). Thus, exploring these cosmological alignments may prove to be useful in order to estimate the contamination from this source in the lensing signal. The importance of angular momentum is also remarkable in semi-analytical modelling of galaxy formation (e.g. \citealt{1998ApJ...505...37A}). The properties of the galactic discs are usually expected to be related to those of the dark matter haloes in which they are embedded (e.g. \citealt{1998MNRAS.295..319M}). However, regarding angular momentum this assumption has been recently questioned \citep{2007MNRAS.380L..58D}.

In this paper, mainly motivated by the observational result found by \citet{2006ApJ...640L.111T}, we use a high-resolution cosmological simulation to study the alignment of shape and angular momentum of galaxy-size dark matter haloes in the shells of large voids. Our simulation provides enough statistics to assess the results in \citet{2006MNRAS.369..335P} and \citet{2007MNRAS.375..184B} and for the first time to measure the strength of the alignment signal of the angular momentum in these haloes without performing a previous selection of them, as opposed to \citet{2007MNRAS.375..184B}.

This work is organized as follows. In Section~\ref{sec:methodology} we describe the cosmological simulation and the methodology. The results are presented in Section~\ref{sec:results}. We summarize and discuss our conclusions in Section~\ref{sec:conclusion}.

\section{MareNostrum Universe simulations and methodology}
\label{sec:methodology}
The cosmological simulation we have used for this study is the \textit{MareNostrum Universe} \citep{2007ApJ...664..117G}. This non-radiative SPH simulation employs $1024^3$ dark matter particles of mass $8.24\times 10^9h^{-1}\rmn{M}_{\sun}$ and $1024^3$ gas particles of mass $1.45\times 10^9h^{-1}\rmn{M}_{\sun}$ in a box of $500 h^{-1}$ Mpc on a side. The cosmological model is $\Lambda$CDM with cosmological parameters $\Omega_{\Lambda}=0.7$, $\Omega_{m}=0.3$, $\Omega_{bar}=0.045$, $h=0.7$, $\sigma_8=0.9$ and a slope $n=1$ for the initial power spectrum. The evolution of the initial conditions was performed by using the TREEPM+SPH code {\sc GADGET-2} \citep{2005MNRAS.364.1105S}. The spatial force resolution was set to an equivalent Plummer gravitational softening of 15 $h^{-1}$ comoving kpc, and the SPH smoothing length was set to the 40$^{th}$ neighbour to each particle. The long range gravitational force calculation was done by the Particle-Mesh algorithm using FFT in an homogeneous mesh of $1024^3$ elements.

In this work we use both dark matter and gas particle distributions of this simulation. In order to test the effects of the baryonic component in the results of our analysis, we have also used the results of an N-body, dark matter only version of the MareNostrum Universe. This simulation has exactly the initial conditions and parameters than the SPH simulation but with $1024^3$ dark matter particles only.

Dark matter haloes are found using a parallel hierarchical friends-of-friends algorithm based on the minimum spanning tree for the particle distribution with linking parameter $b=0.17$, which corresponds to overdensity $\Delta=330$ with respect to mean matter density. Using equation 2 in \citet{2005ApJ...627..647B}, and taking into account that the mean axis ratios for galaxy-size haloes are $c/b\simeq 0.79$ and $b/a\simeq 0.75$, we find that about 65 particles are enough to determine main axes orientations to within $15\degr$. As this is a reasonable resolution, we are restricted to use haloes with $M > 5.4\times 10^{11} h^{-1}\rmn{M}_{\sun}$, so we will choose our lower mass limit above this value. There are 881,861 haloes in the entire simulation box above this mass at $z=0$, which provides very good statistics for our purposes.

The detection of voids is made using the {\sc HB} void finder \citep{2006MNRAS.369..335P}. This code searches for the maximal non-overlapping spheres which are larger than a given radius and empty of objects below a given mass. Those spheres are our voids. We are interested only in large voids, i.e. those with $R_{\rmn{void}} > 10h^{-1}$ Mpc, which characterize the orientation of the large-scale distribution at the position of the halo. We searched for voids defined by haloes with masses larger than $7.95\times 10^{11}h^{-1}\rmn{M}_{\sun}$, whose axes are resolved to within $13\degr$. The mass limit has been chosen to get the same number density of haloes as the number density of galaxies found in the observed galaxy samples of \citet{2006ApJ...640L.111T} ($n=5\times 10^{-3}h^3$ Mpc$^{-3}$, which corresponds to galaxies brighter than $M_{b_J} \geq -19.4 +5\log h$). With this defining mass we found a total of 3047 voids, with a median radius of $11.17h^{-1}$ Mpc. For further comparison with the observational result, we look for Milky Way-sized haloes in a shell of $4h^{-1}$ Mpc around the surface of each void. We thus remove all the haloes with mass below the defining mass of the voids $7.95\times 10^{11}h^{-1}\rmn{M}_{\sun}$ (which is greater than our resolution limit), and those with mass exceeding $7.95\times 10^{12}h^{-1}\rmn{M}_{\sun}$, so that they are still galaxy-size. Considering all these constraints, we are finally left with 88,426 haloes in the shells of voids and in this range of mass for our analysis.

For every halo we calculate its total angular momentum,
\begin{equation}
\bmath{L}=\sum_k m\bmath{r}_k\bmath{\times}\bmath{v}_k
\end{equation}
where $\bmath{r}_k$ has its origin in the centre of mass of the halo and the sum is over every dark matter particle identified by the halo finder as belonging to the halo. The orientation of the angular momentum with the centre of the void is given by the $\theta$ angle derived from:
\begin{equation}
\cos\theta=\frac{\bmath{R}\bmath{\cdot}\bmath{L}}{|\bmath{R}||\bmath{L}|}
\label{eqn:calc}
\end{equation}
where $\bmath{R}$ is the vector linking the centre of mass of the halo with the centre of the void, $\bmath{R}=\bmath{r}_{\rmn{halo}}-\bmath{r}_{\rmn{void}}$. In the same way we use an analogous equation to calculate the angle between $\bmath{R}$ and the major ($\bmath{a}$), middle ($\bmath{b}$) and minor ($\bmath{c}$) halo axis, i.e. the three main axes resulting of the diagonalization of the 'inertia' tensor of haloes in the shells of the voids (the eigenvalues will be referred to as $a$, $b$ and $c$). This tensor is defined as:
\begin{equation}
\mathbfss{I}_{ij}=\sum_k x_{i,k} x_{j,k}
\end{equation}
where the coordinates $x_i,k$ represent the $i$-th component of the position vector of the $k$-th particle, measured with respect to the centre of mass of the halo. This sum is again over halo particles (see \citealt{2006ApJ...652L..75P}).

\section{Results}
\label{sec:results}
\begin{figure*}
\includegraphics[width=0.9\textwidth]{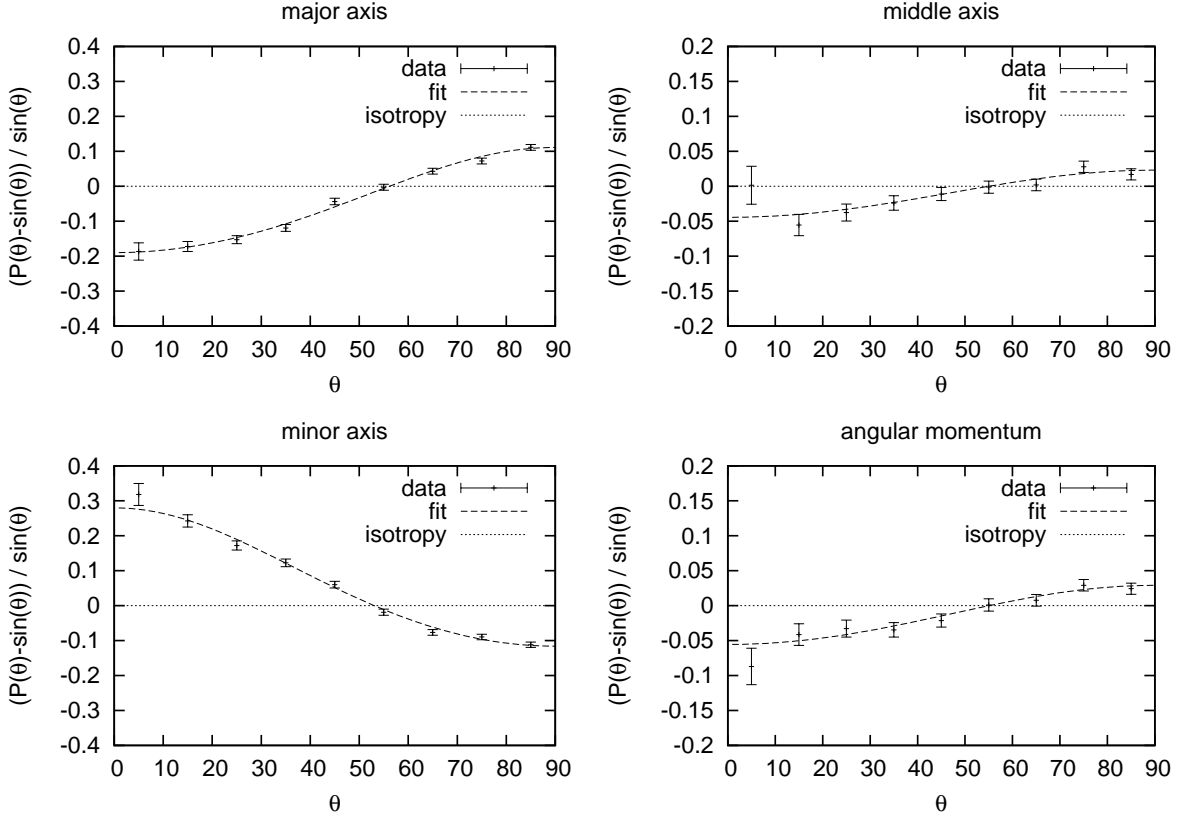}
\caption{Probability density distribution (measured with respect to the case of isotropic orientations) for the angles between the angular momentum of dark matter haloes in the shells of voids, the three principal axes of these haloes, and the direction to the centre of the void. The error bar on each bin represents the Poissonian error. The dashed line shows a fit of the data using Eq.~\ref{eqn:fit}, and the dotted line represents the null hypothesis. Taking haloes in the shells of spheres with randomly distributed centres makes the results agree with isotropy.}
\label{fig:main}
\end{figure*}

The main results are summarized in Fig.~\ref{fig:main} and Table~\ref{tab:axes}. We show in Fig.~\ref{fig:main} the probability density distribution of the orientations of the three main axes and angular momentum of haloes in the shells of voids (using a spherical shell of $4h^{-1}$ Mpc around the surface of the void), with respect to the direction to the centre of the void. The angles are calculated using equation~\ref{eqn:calc} and the analogous ones for the main axes correspondingly.

We also plot an analytical fit of the probability density distribution (Betancort-Rijo \& Trujillo, in preparation) given by the following equation:
\begin{equation}
P(\mu)\rmn{d}\mu\propto\frac{p\rmn{d}\mu}{\left(1+\left(p^2-1\right)\mu^2\right)^{3/2}} ; \quad \mu\equiv\cos\theta
\label{eqn:fit}
\end{equation}
where $p$ is a free parameter, related to the ratio between the dispersion in the radial component of the angular momentum vector, and its dispersion in the transverse component. Although this equation was proposed to describe how the orientation of angular momentum is distributed, it proves to be a very good fit for the distribution of the orientations of the halo main axes.

The distribution of an isotropic distribution is computed and also shown in Fig.~\ref{fig:main}. If these vectors have no particular orientation (our null hypothesis), then the expected probability distribution is a sine function. We use this fact to represent the deviation of the probability distributions with respect to the isotropic case, so that the null hypothesis corresponds to the zero level in this plot.

In order to quantify the strength of these alignments, we use the parameter $p$ in Eq.~\ref{eqn:fit}, as previously done by \citet{2007MNRAS.375..184B}. The null hypothesis corresponds to $p=1$, so that $\mu$ is uniformly distributed. Lower and higher values of $p$ correspond to a preferential orientation parallel and perpendicular to the centre of the void respectively. A good approximation for this parameter in the case $p\simeq1$ comes from the relation:
\begin{equation}
\left\langle\cos\theta\right\rangle =\frac{1}{1+p}
\label{eqn:estim}
\end{equation}
where the average is taken over all the halo sample in the shells of voids.

The results presented in Table~\ref{tab:axes} are for the three main axes and angular momentum of haloes in the shell of $4h^{-1}$ Mpc around the surface of the void. The first column shows the result of averaging the cosine values of the angles considered here for our sample of haloes. The second column shows the result of the statistical test performed here, i.e. the average of the cosine test. In the case of isotropic orientations, the average of the cosine is normally distributed with mean 0.5 and variance $1/\left(12N\right)$, where $N$ is the number of haloes. Therefore, the bigger is the departure from the mean and the bigger the statistics, the higher the significance of rejecting the null hypothesis, $N_{\sigma}=\sqrt{12N}(\left< \cos\theta \right>-0.5)$. The third column represents the strength of the alignment using the aforementioned parametrization, estimated from a Levenberg-Marquardt fit to our data using Eq.~\ref{eqn:fit}. The fourth column also represents the value of $p$ but this time it is estimated using Eq.~\ref{eqn:estim}. The error bars in this parameter correspond to the 68\% confidence level. The last column shows that the value of the reduced $\chi^2$ is indeed near to unity, which proves the goodness of the fit. It is important to remark that the statistical test only quantifies the probability of isotropy in these orientations; the magnitude of the alignment is instead measured by the parameter $p$.

\begin{table*}
\caption{The strength of the alignment of the three main axes and angular momentum vector of haloes (and baryons embedded therein) in a shell of $4h^{-1}$ Mpc around large voids, with respect to the direction to the centre of the void. First column shows the average of the cosine of the angles between these vectors and the radial direction $\bmath{R}$. Second and third columns show the value of the parameter $p$ from a fit to Eq.~\ref{eqn:fit} and using Eq.~\ref{eqn:estim} respectively. The fourth column shows the goodness of this fit. The last column represents the result of the statistical test for the rejection of the null hypothesis in terms of $N_{\sigma}$ (see text for details).}
\begin{tabular}{|cccccr@{.}l|}
\hline
axis &  $\left< \cos\theta \right>$ & $p_{\rmn{fit}}$ & $p_{\rmn{estimated}}$ & $\frac{\chi^2-N_{\rmn{d.o.f.}}}{\sqrt{2N_{\rmn{d.o.f.}}}}$ & \multicolumn{2}{c}{$N_{\sigma}$} \\
\hline
\multicolumn{7}{c}{dark matter} \\
\hline
major  & 0.4739 & $1.110 \pm 0.004$ & $1.110 \pm 0.004$ & +0.06 & -26&9 \\
minor  & 0.5306 & $0.884 \pm 0.003$ & $0.885 \pm 0.003$ & +0.63 & +31&6 \\
middle & 0.4945 & $1.023 \pm 0.004$ & $1.022 \pm 0.004$ & -0.04 & -5&7 \\
$\bmath{L}$ & 0.4926 & $1.029 \pm 0.004$ & $1.030 \pm 0.004$ & -0.69 & -7&6 \\
\hline
\multicolumn{7}{c}{gas} \\
\hline
major  & 0.4639 & $1.154 \pm 0.012$ & $1.156 \pm 0.013$ & -0.56 & -13&1 \\
minor  & 0.5288 & $0.891 \pm 0.009$ & $0.891 \pm 0.010$ & -0.36 & +10&4 \\
middle & 0.5042 & $0.980 \pm 0.010$ & $0.983 \pm 0.011$ & -1.15 & +1&5 \\
$\bmath{L}$ & 0.4922 & $1.030 \pm 0.011$ & $1.032 \pm 0.011$ & +3.20 & -2&8 \\
\hline
\end{tabular}
\label{tab:axes}
\end{table*}

We find for this sample of haloes that the minor axis is preferentially aligned with the direction to the centre of the void, and the major axis is preferentially aligned perpendicular to it. Middle axis also shows an alignment orthogonal to the radial direction, but much weaker than previous one. This confirms the results reported by \citet{2006ApJ...652L..75P} and \citet{2007MNRAS.375..184B}. The same conclusion holds from the analysis of the gas particles embedded in dark matter haloes, although in this case middle axis does not show any significant alignment.

In order to characterize the complete statistical information about the spatial orientation of these haloes, we studied the distribution of the Euler angles. These angles link two orthogonal systems defined at location of each halo in the shell of a void: the one formed by the direction to the centre of the void and any two perpendicular directions that lie in the shell $\left(\bmath{X},\bmath{Y},\bmath{R}\right)$, and the one formed by the three main axes derived from the diagonalization of the inertia tensor of the halo $\left(\bmath{a},\bmath{b},\bmath{c}\right)$. The second coordinate system can be obtained from the first by means of a set of three rotations: a rotation of angle $\phi$ about $\bmath{R}$, a rotation of angle $\omega$ about the rotated $\bmath{X}$ axis (which places the $\bmath{R}$ axis in the direction of $\bmath{a}$), and finally a rotation of angle $\psi$ about this $\bmath{a}$ axis, which places the other two in the direction of $\bmath{b}$ and $\bmath{c}$. It is important to remark that $\omega$ is actually the angle between the major axis and the direction to the centre of the void, so we already know its distribution. Moreover, due to the rotational symmetry about the direction $\bmath{R}$ (there is no privileged point in the surface of a sphere), we also know that $\phi$ is uniformly distributed. Thus, for a complete description of the spatial orientation of our haloes we need to find out the distribution of the $\psi$ angle. Obviously if the shape of the haloes were not aligned with any particular direction, $\psi$ would be uniformly distributed. Nevertheless we find that the distribution of this angle is instead well described by (Betancort-Rijo \& Trujillo, in preparation)
\begin{equation}
P\left(\psi\right)\rmn{d}\psi=\frac{2}{\pi}\left(1+\beta\cos\left(2\psi\right)\right)\rmn{d}\psi
\end{equation}
where $\beta$ is a free parameter. A fit of our data gives $\beta=-0.093\pm0.005$. However, if we select a sample of those haloes whose major axis forms an angle with $\bmath{R}$ above a given value $\omega_0$, this parameter seems to rise. This is an indication that in fact, the distribution of $\psi$ is not independent of this angle, and the complete description of the shape alignment will be given by a more general distribution $P(\omega,\psi)$.

But our most remarkable result is that we find a previously undetected alignment of angular momentum, whose orientation is preferentially perpendicular to the direction to the centre of the void (but not necessarily aligned with the major axis). This is confirmed by the average of the cosine and a Kolmogorov--Smirnov test. Although this alignment is small, both statistical tests agree in rejecting the null-hypothesis with a probability corresponding to a significance level of more than $7\sigma$.

Previous studies have not found such an orientation of the angular momentum in the shells of voids, mainly because of the poor halo statistics due to the small simulation box. For instance, the halo sample in \citet{2006ApJ...652L..75P} consisted of only 1729 objects with which they managed to find $p=0.96\pm0.04$. Despite this value is compatible with our result, it points that angular momentum tends to be aligned with the direction to the centre of the void, which is the opposite of what we find. Box size may also have been the reason why \citet{2006MNRAS.371..750H} could not reject the null hypothesis (causing the error bars to be about a factor of two larger than those shown in our Fig.~\ref{fig:main}), although other factors may have had some influence on their conclusions, like the small number of particles in their haloes or their estimation of error bars via bootstrap method. The case of \citet{2007MNRAS.375..184B} is specially surprising, since having the same box size, the resolution of Millennium simulation largely exceeds that of this paper. Yet, they still find that the distribution of the angular momentum is compatible with a random orientation. Taking a convenient subsample of haloes they found a significant alignment, but only in one out of the seven radial bins considered. This subsample is obtained selecting haloes with a disc-dominated galaxy at their centre, using semianalytic galaxy catalogue of \citet{2006MNRAS.365...11C}. The galaxy must be brighter than $M_K < -23$ and have a bulge to total ratio $0 < B/T < 0.4$. On the contrary, it is noteworthy that no preselection was performed for the results presented here.

The likely explanation for this disagreement is that their haloes were identified with the same halo finder but with a linking length which corresponds to an overdensity 900 times the mean matter density, to focus on the core properties of the haloes. Here, we compute main axes and angular momentum of haloes detected with a linking length corresponding to 330 times the mean matter density (see Section~\ref{sec:methodology}). For comparison, we show in Table~\ref{tab:over} the results of the same analysis as in Table~\ref{tab:axes} but using a catalogue of haloes in our simulation box defined by an overdensity eight times higher. Although the main axes show a stronger alignment in the inner parts of the halo, it is clear that the alignment of the angular momentum with the direction to the centre of the void is lost when we focus in the inner parts. The contribution to the angular momentum from the particles in the outer shells of a halo is then fundamental for this conclusion. It is quite remarkable that the gas in the inner part does not show this alignment either, despite the fact that both major and minor axes show an enhanced alignment compared to dark matter haloes.

\begin{table*}
\caption{Same as Table~\ref{tab:axes}, but the haloes are defined here as enclosing eight times the virial overdensity.}
\begin{tabular}{|cccccr@{.}l|}
\hline
axis &  $\left< \cos\theta \right>$ & $p_{\rmn{fit}}$ & $p_{\rmn{estimated}}$ & $\frac{\chi^2-N_{\rmn{d.o.f.}}}{\sqrt{2N_{\rmn{d.o.f.}}}}$ & \multicolumn{2}{c}{$N_{\sigma}$} \\
\hline
\multicolumn{7}{c}{dark matter} \\
\hline
major  & 0.4702 & $1.125 \pm 0.005$ & $1.127 \pm 0.005$ & -0.23 & -25&2 \\
minor  & 0.5362 & $0.869 \pm 0.004$ & $0.865 \pm 0.004$ & +1.00 & +30&7 \\
middle & 0.4925 & $1.030 \pm 0.005$ & $1.030 \pm 0.005$ & -1.42 & -6&3 \\
$\bmath{L}$ & 0.5003 & $1.000 \pm 0.005$ & $0.999 \pm 0.005$ & -0.70 & 0&2 \\
\hline
\multicolumn{7}{c}{gas} \\
\hline
major  & 0.4476 & $1.232 \pm 0.020$ & $1.234 \pm 0.021$ & -0.65 & -12&3 \\
minor  & 0.5549 & $0.793 \pm 0.013$ & $0.802 \pm 0.014$ & +2.54 & +12&8 \\
middle & 0.4926 & $1.029 \pm 0.017$ & $1.030 \pm 0.018$ & +0.95 & -1&7 \\
$\bmath{L}$ & 0.5015 & $0.988 \pm 0.016$ & $0.994 \pm 0.017$ & -0.19 & +0&4 \\
\hline
\end{tabular}
\label{tab:over}
\end{table*}

As a further test of our results we repeated the analysis to obtain the strength of the alignments, by taking haloes in the shells of \textit{fake} voids (i.e. spheres with randomly distributed centres throughout the simulation box). The results for both shape and angular momentum alignment are then compatible with the null hypothesis. To test the effect of baryons in the alignments of dark matter haloes we have analysed the dark matter only version of the MareNostrum Universe. The results we obtain for the strength of these alignments are compatible to those listed in the upper (dark matter) part of Table~\ref{tab:axes} and Table~\ref{tab:over}.

We also studied the dependence of the strength of these alignments when moving outside the surface of the void. The results are shown in Table~\ref{tab:strength}. In this table we show the value of the parameter $p$ estimated using Eq.~\ref{eqn:fit} along with their error bars, in several different shells surrounding the void. The shells are selected for direct comparison with \citet{2007MNRAS.375..184B}. It is clear from the table that the strength of the alignments dilutes as we move outside the surface of the void. The inner shell show a weaker alignment of the angular momentum compared to that work. Instead, this strength declines slower here as we advance to the outer shells. This is in contradiction with the statement in \citet{2007MNRAS.375..184B} suggesting that taking a wide shell can mask this alignment. Interestingly, we find that the alignment of the angular momentum is still significant even at few times $R_{\rmn{void}}$.

\begin{table*}
\caption{The strength of dark matter halo alignments measured from parameter $p$ at 1$\sigma$ confidence level using different shells around the surface of the void. Recall that the null hypothesis (isotropic orientations) corresponds to $p=1$. Lower and higher values of $p$ correspond to a preferential orientation parallel and perpendicular to the centre of the void respectively.} \begin{tabular}{|cccccc|}
\hline
Shell & Major axis & Middle axis & Minor axis & Angular momentum & Number of \\
$R_{\rmn{void}}$ Units & $p$ & $p$ & $p$ & $p$ & haloes \\
\hline
$1.00 < R < 1.05$ & $1.138 \pm 0.010$ & $1.022 \pm 0.009$ & $0.865 \pm 0.007$ & $1.046 \pm 0.009$ & 16406\\
$1.05 < R < 1.10$ & $1.149 \pm 0.013$ & $1.008 \pm 0.011$ & $0.874 \pm 0.010$ & $1.024 \pm 0.012$ & 9437\\
$1.10 < R < 1.20$ & $1.115 \pm 0.008$ & $1.019 \pm 0.007$ & $0.883 \pm 0.006$ & $1.021 \pm 0.007$ & 24329\\
$1.20 < R < 1.40$ & $1.085 \pm 0.005$ & $1.027 \pm 0.004$ & $0.901 \pm 0.004$ & $1.024 \pm 0.004$ & 66576\\
$1.40 < R < 1.80$ & $1.052 \pm 0.003$ & $1.006 \pm 0.003$ & $0.946 \pm 0.002$ & $1.013 \pm 0.003$ & 187565\\
$1.80 < R < 2.60$ & $1.025 \pm 0.002$ & $1.008 \pm 0.002$ & $0.969 \pm 0.002$ & $1.009 \pm 0.002$ & 460592\\
$2.60 < R < 3.20$ & $1.013 \pm 0.001$ & $1.001 \pm 0.001$ & $0.985 \pm 0.001$ & $1.006 \pm 0.001$ & 555621\\
\hline
\end{tabular}
\label{tab:strength}
\end{table*}

\begin{figure}
\includegraphics[width=0.5\textwidth]{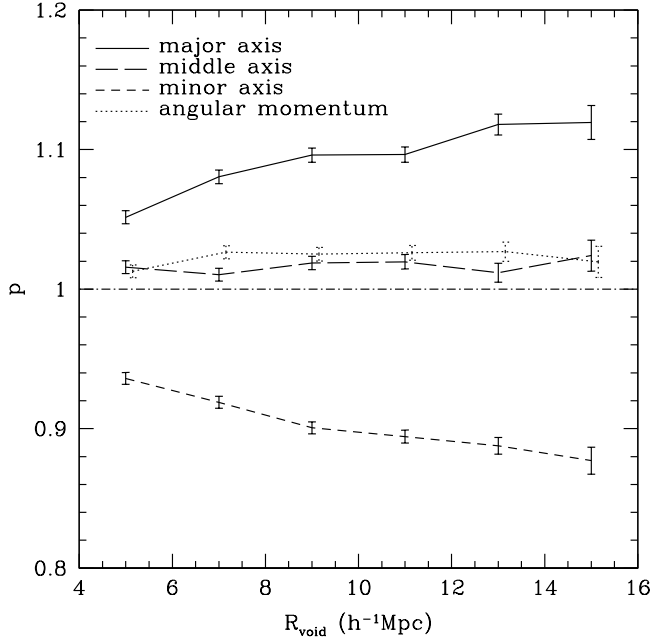}
\caption{The strength of the alignment as a function of the void size, measured over bins of $2h^{-1}$ Mpc wide in void radius. The solid, long-dashed and short-dashed lines represent the major, middle and minor axes respectively. Dotted line represents the angular momentum, which has been slightly shifted to the right for the sake of clarity. Error bars show the 1$\sigma$ uncertainty in the value of the parameter $p$. The dash-dotted line represents the null hypothesis.}
\label{fig:size}
\end{figure}

\begin{figure}
\includegraphics[width=0.5\textwidth]{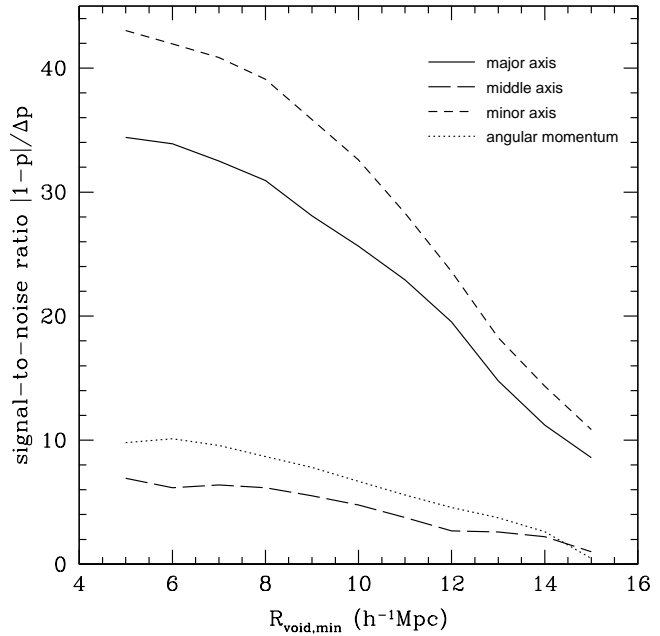}
\caption{The signal-to-noise ratio in the alignments of haloes around voids as a function of the minimum void size. This ratio is estimated by the deviation of the parameter $p$ from the isotropic case $p=1$, over the uncertainty in this parameter. As in Fig.~\ref{fig:size}, dotted line represents the angular momentum and the solid, long-dashed and short-dashed lines represent the major, middle and minor axes respectively.}
\label{fig:signal}
\end{figure}

The minimum void size $R_{\rmn{void}} > 10h^{-1}$ Mpc was chosen in order to mimic the observational result of \citet{2006ApJ...640L.111T}, and it seems reasonable as it is of the order of the scale of nonlinearity at present, and hence haloes around voids with this size are embedded into quasi-linear structures which remember the original shape. However, we are in an excellent position to study whether this value is good enough for our analysis. In Fig.~\ref{fig:size} we show the dependence of the strength of the alignments on the choice of the void size, in the range of $5h^{-1} \rmn{Mpc} < R_{\rmn{void}} < 15h^{-1} \rmn{Mpc}$. The width of the bins is chosen to be $\Delta R_{\rmn{void}} = 2h^{-1}$ Mpc. All the haloes in a shell $0.4 R_{\rmn{void}}$ in width around the surface of the void are taken into account. We find that the alignment decreases considerably with the radius of the void, fading away for $R_{\rmn{void}} < 5h^{-1}$ Mpc. On the other hand, large voids are rare and the poor statistics of objects around them increases the uncertainty on the parameter $p$. Therefore, a strong and significant alignment can be found using voids of intermediate size. It is worthy noting that if we use the same procedure as above, i.e. all the voids larger than a given minimum size are included into the analysis, the highest signal-to-noise ratio is located around $\lesssim 6h^{-1}$ Mpc. As shown in Fig.~\ref{fig:signal}, this ratio declines very steeply with the choice of minimum void size because the decrease in the statistics (and hence the increase in the uncertainty $\Delta p$) is much higher than the increase in the strength of the alignment. This result is of interest in observational studies, but it is important to remark that the position of this maximum could be shifted when using redshift space instead of real space.

Some studies have suggested that voids are not generally spherical, so that alignment detections assuming spherical voids should therefore be seen as residual of genuine physical alignments (e.g. \citealt{2007ApJ...655L...5A}). However, using an ellipsoidal void catalogue for the MareNostrum Universe we find similar results for the orientation of angular momentum and main axes for the haloes in the shells of those non-spherical voids (see Table~\ref{tab:ellip}). Moreover, we find no significant difference between the direction to the centre of the void and the direction perpendicular to the ellipsoid, taking $\bmath{n}\equiv\left(\bmath{R}\bmath{\cdot}\bmath{a}/a^2,\bmath{R}\bmath{\cdot}\bmath{b}/b^2 ,\bmath{R}\bmath{\cdot}\bmath{c}/c^2\right)$ as the reference direction for angle measurements. Dark matter haloes selected here are located in a shell of $4h^{-1}$ Mpc thick around the maximal sphere enclosed into the ellipsoid. In order to study more in depth the relevance of void shape, we studied the probability distribution function of the Euler angle $\phi$. In the case of spherical voids, this angle is uniformly distributed due to the rotational symmetry in the plane tangential to the sphere at the position of the halo. On the other hand, this symmetry is broken in the case of ellipsoidal voids, so the distribution of the $\phi$ angle might be different. However, taking into account that
\begin{equation}
\sin\phi=\frac{\bmath{v \cdot i} -\left(\bmath{v \cdot R}\right)\left(\bmath{i \cdot R}\right)}{\sqrt{1-\left(\bmath{v \cdot R}\right)^2}\sqrt{1-\left(\bmath{i \cdot R}\right)^2}}
\end{equation}
where $\bmath{i}$ is defined in the direction of the major axis of the void and $\bmath{v}$ in the direction of the major axis of the halo, we find $\left\langle\phi\right\rangle/\left(\pi/2\right)=0.5001\pm 0.0009$ in agreement with a uniform distribution.

\begin{table}
\caption{Values for the $p$ parameter for the alignment of the three main axes and angular momentum vector with the direction to the centre of the void for haloes in a shell of $4h^{-1}$ Mpc using two different void catalogues in the same simulation. For the ellipsoidal void catalogue the results for both the \textit{radial} and the \textit{normal} direction are shown.}
\begin{tabular}{|cccc|}
\hline
axis & spherical void catalogue & \multicolumn{2}{c}{ellipsoidal void catalogue}\\
 & $p$ (using $\bmath{R}\equiv\bmath{n}$) & $p$ (using $\bmath{R}$) & $p$ (using $\bmath{n}$) \\
\hline
major  & $1.110 \pm 0.004$ & $1.112 \pm 0.004$ &$1.112 \pm 0.004$\\
minor  & $0.884 \pm 0.003$ & $0.886 \pm 0.004$ &$0.887 \pm 0.003$\\
middle & $1.023 \pm 0.004$ & $1.021 \pm 0.004$ &$1.021 \pm 0.004$\\
$\bmath{L}$ & $1.029 \pm 0.004$ & $1.028 \pm 0.004$ &$1.029 \pm 0.004$\\
\hline
\end{tabular}
\label{tab:ellip}
\end{table}

There also have been recent claims of dependence on the alignment of the angular momentum with the mass of the halo in filaments and walls of the large-scale structure \citep{2007ApJ...655L...5A,2007MNRAS.375..489H}. In agreement with these papers we find a slight trend of increasing alignment strength with halo mass for our sample. Using a shell of $4h^{-1}$ Mpc the haloes near the resolution limit imposed, with $M= 5\times10^{11}$--$8\times10^{11}h^{-1}\rmn{M}_{\sun}$, have a strength value of $p\simeq 1.015$. On the other hand, haloes with masses in the upper limit of the sample $M= 8\times10^{12}$--$2\times10^{13}h^{-1}\rmn{M}_{\sun}$ show evidence of a stronger alignment with $p\simeq 1.060$.

\section{Discussion and conclusion}
\label{sec:conclusion}
In this paper we have analysed the orientation of main axes and angular momentum of dark matter haloes in the shells of large voids using the MareNostrum Universe simulation. The spins of these haloes are preferentially perpendicular to the direction to the centre of the void. The possibility of isotropic orientations has been rejected at $7\sigma$ confidence level. This result is in qualitative agreement with and is likely to be related to the observational result of \citet{2006ApJ...640L.111T} for spiral galaxies using large galaxy surveys. The alignment of dark matter collapsed structures with the large-scale structure finds a natural explanation in the context of the Tidal Torque Theory. The different components of the angular momentum vector present a dispersion which is not isotropic: there is a marked asymmetry between the dispersion in the direction of maximum compression as compared to that in the direction of maximum expansion (Betancort-Rijo \& Trujillo, in preparation). Therefore, very low density regions such as large voids are suitable to study this effect as they are surrounded by a region with a large tidal field, causing the anisotropy in the shear tensor we are measuring in this work. It is worthy noting that this effect is not due to accretion of matter along filaments. Large voids are surrounded by a complex filamentary network of matter \citep{2005MNRAS.359..272C} which could represent a preferential direction of infall (but see \citealt{2002ApJ...581..799V}). This accretion generates angular momentum pointing perpendicular to the filament in which the haloes are embedded, which could be either the direction to the centre of the void but could also be perpendicular to it, resulting in no preferential orientation of angular momentum. Therefore, the anisotropic dispersion in the angular momentum vector seems more plausible in order to explain of this effect.

This is the first time that this alignment has been found in cosmological simulations using haloes in the shells of voids with no preselection of the haloes, in contrast to previous results. The poor halo statistics in the simulation box, the small number of particles in haloes, and even an overestimation of the error bars are likely to be the causes of the non-detection of this effect in \citet{2006ApJ...652L..75P} and \citet{2006MNRAS.371..750H}. \citet{2007MNRAS.375..184B} did not succeed in finding this alignment either, and concluded that the distribution of angular momentum vectors is compatible with isotropy for their total dark matter halo sample. This is somewhat surprising since they used the Millennium simulation which exceeds the resolution of the MareNostrum Universe simulation used here, having both the same box size. However, we have checked that the uncertainties in the orientation of the vectors used here are small enough to ensure the veracity of these results. We recall that the haloes used here are chosen to be well resolved and their directions are determined to within an angle of $13\degr$. Indeed, it can be proven that this small indetermination in the angle measurement makes the uncertainty in the parameter $p$ be negligible. It is very likely that the linking length (and hence the overdensity) used to identify the haloes in the simulation box makes the results to differ as we have shown in Section~\ref{sec:results}. The choice of an overdensity 900 times the mean matter density (as opposed to our selection of overdensity 330) allowed them to focus only in the inner parts of the halo, but this may have prevented them from the detection of the alignment of angular momentum we find. This suggests that the particles in the outer shells of the halo are determining in the measurement of the strength of this alignment.

None the less, the claim of \citet{2007MNRAS.375..184B} that they found a significant alignment at the surface of the void, is based only in the choice of a subsample based on a semianalytic galaxy catalogue. As we have mentioned above, they do not find any preferential orientation in their total sample. Moreover, we find that this alignment is significant even at few times $R_{\rmn{void}}$. The signal decreases with distance to the void centre, which is expected for several reasons: the dispersion in the different components of the angular momentum vector becomes isotropic with increasing distance to the centre of the void. This is due to the less anisotropic tidal field in regions with higher density. Besides, the local density increases at larger distance from the centre of the void, and hence the interaction with neighbouring haloes is stronger, diluting the alignment from the initial torquing.

The choice of the minimum size of the void $R_{\rmn{void}}>10h^{-1}$ Mpc has proven to be reasonable because it is a good compromise between strength of the alignment and statistics. However, we have found that the choice $R_{\rmn{void}}>6h^{-1}$ Mpc is even better for the detection of alignment of dark matter haloes. When including smaller voids, which are more abundant, the statistics of haloes in the sample is highly improved. On the other hand, the alignment of these haloes is much weaker compared to the strength of the haloes around the largest voids: haloes around smaller voids are separated by shorter distances, and thus they are more influenced by interactions. Thus, the best choice is non-trivial and requires from the large amount of data given by cosmological simulations. Prospective observational studies should find this scale useful to get high signal-to-noise ratios in the detection of these alignments.

Our main result that galaxy-size dark matter haloes around voids have spins that lie in the shells of voids, is in agreement with that of \citet{2007ApJ...655L...5A} if haloes in the shells of void which form large-scale sheets are mainly responsible for this effect. However, \citet{2007MNRAS.381...41H} found a stronger alignment of haloes in sheets than largely exceed the one presented here. Nevertheless, haloes around voids are located in a less dense environment than actual sheet haloes, which may have influence on their orientation. 

Shape alignment of haloes in the surface of large voids is further confirmed, as shown previously by \citet{2006ApJ...652L..75P}. Minor axis is found to be preferentially aligned with the direction to the centre of the void and major axis tends to be aligned with the orthogonal direction. These results are also robust with respect to the shape of the void. Using an ellipsoidal void catalogue and measuring angles with respect to the radial and the normal direction we do not find any difference compared to the results obtained using spherical voids for the orientation of the main axes and angular momentum. Hence, we do not find that the shape of the void is meaningful for this purpose. We also analysed the strength of these alignments in several shells around the surface of the void. Comparing our results with those in \citet{2007MNRAS.375..184B} we find a systematically weaker alignment in the shape orientation of dark matter haloes.

Regarding gas particles, we find that the angular momentum of baryons is oriented with respect to the radial direction with a strength similar to dark matter, even considering the inner parts of the halo. However, in the core of the halo shape alignment for gas particles is enhanced with respect to dark matter. Although this is a remarkable result for its observational implications, the effect of cooling and full scale star formation in these orientations remains to be investigated.

In addition, we found a slight trend of the alignment of angular momentum with the halo mass, as in \citet{2007ApJ...655L...5A}. More massive haloes happen to be more aligned with the orthogonal direction. A possible explanation to this fact could be their recent formation, as pointed out by  \citet{2007ApJ...655L...5A}, which makes less massive haloes be more affected by non-linear interactions. It has also been suggested that an alternative explanation is that this effect might be due to the preferred accretion along the filaments \citep{2006ApJ...646..815S}.

Evolution of this alignment in cosmological simulations have been studied by \citet{2007MNRAS.381...41H}. Unfortunately, it is very difficult to draw a conclusion about the evolution of the alignments from the present data. A deeper analysis of the redshift evolution of the strength of this signal is mandatory to elucidate its origin.\\

We thank Barcelona Supercomputer Center - Centro Nacional de Supercomputaci\'on for the computer time awarded to run the MareNostrum Universe simulations. The analysis of this simulation has been performed in NIC J\"ulich. A.J.C. enjoyed the hospitality of the Astrophysical Institute Potsdam where part of this work was carried out. A.J.C., J.B.-R., S.G.P. and F.P. thank the Spanish MEC under grant PNAYA 2005-07789 for their support. S.G. acknowledges the support of the European Science Foundation through the ASTROSIM Exchange Visits Programme. G.Y. acknowledges financial support to the Spanish PNFPA2006-01105 and PNAYA2006-15492-C03. A.J.C. appreciates the financial support of the MEC through Spanish grant FPU AP2005-1826. This work was also partially supported by the Acciones Integradas Hispano-Alemanas.

\bibliography{mycites}

\bsp
\end{document}